\documentclass[aps,prb,twocolumn]{revtex4-1}
\usepackage{gensymb}
\usepackage{soul}
\usepackage{color}\label{key}
\usepackage{graphicx}
\usepackage{epstopdf}
\usepackage{adjustbox}
\usepackage{array}
\newcolumntype{P}[1]{>{\centering\arraybackslash}p{#1}}
\usepackage{amsmath}  
\begin{document}
\title{An experimental and Ab-initio study of Electronic and Magnetic properties of FeGa$_3$}
\author{Debashis Mondal$^{1,2*}$, Soma Banik$^1$, C. Kamal$^3$, Parasmani Rajput$^4$, Arumugam Thamizhavel$^5$, A. Banerjee$^6$, Aparna Chakrabarti$^{2,3}$, Tapas Ganguli$^{1,2}$}
\address{$^1$Synchrotrons Utilization Section, Raja Ramanna Centre for Advanced Technology, Indore, 452013, India.}
\address{$^2$Homi Bhabha National Institute, Training School Complex, Anushakti Nagar, Mumbai, 400094, India.}
\address{$^3$Theory and Simulations Lab, HRDS, Raja Ramanna Centre for Advanced Technology, Indore, 452013, India.}
\address{$^4$Atomic $\&$ Molecular Physics Division, Bhabha Atomic Research Centere, Trombay, Mumbai-400085, India.}
\address{$^5$Tata Institute of Fundamental Research, Homi Bhabha Road, Colaba, Mumbai 400-005, India.}
\address{$^6$UGC-DAE Consortium for Scientific Research, University Campus, Khandwa Road, Indore, 452001,India.}
\email{debashis@rrcat.gov.in, tapas@rrcat.gov.in}

\begin{abstract}
Electronic structure of FeGa$_3$ has been studied using experiments and ab-initio calculations. Magnetization measurements show that FeGa$_3$ is inherently diamagnetic in nature. Our studies indicate that the previously reported magnetic moment on the Fe atoms in FeGa$_3$ is not an intrinsic property of FeGa$_3$, but is primarily due to the presence of disorder, defects, grain boundaries etc that break the symmetry about the Fe dimers. Analysis of the results obtained from magnetic measurements, photoelectron spectroscopy, Fe K-edge X-ray absorption near edge spectroscopy and ab-initio calculations clearly indicates that, the effects of on-site Coulomb repulsion between the Fe 3d electrons do not play any role in determining the electronic and magnetic properties of FeGa$_3$. Detailed analysis of results of single crystal and poycrystalline FeGa$_3$, helps to resolve the discrepancy in the electronic and magnetic properties in FeGa$_3$ existing in the literature, consistently. 
\end{abstract}

\maketitle

\section{Introduction}

Intermetallic materials are very interesting and technologically important because of their electronic, magnetic and mechanical properties. Although most of the intermetallics are metallic in nature, a few of these systems like FeSi, FeSb$_2$, RuAl$_2$, FeGa$_3$, RuGa$_3$ and RuIn$_3$ \cite{Jaccarino, Petrovic,Weinert,Amagai,Bogdanov} show semiconducting behavior with a small bandgap. In all these materials, the hybridization between the d orbitals of the transition metal and p orbitals of p block element splits the density of states at the Fermi level, thereby creating a bandgap. \cite{Ulrich2} In this set of $nd^6$ based intermetallics, FeSi and FeSb$_2$ have a small gap of approximately 0.06 and 0.04 eV respectively, \cite{Jaccarino, Petrovic} whereas, a few other materials such as FeGa$_3$, RuGa$_3$, OsGa$_3$, RuIn$_3$ and RuAl$_2$ have a relatively higher bandgap of 0.26, 0.32, 0.42, 0.4 - 0.5 and 0.6 eV respectively. \cite{Amagai,Bogdanov, Mandrus} The bandgap in most of these systems lie in the IR region of the EM spectra, thereby opening the possibility of using these materials for IR devices. The narrow bandgap with a large density of states near the valence band edge in these intermetallics, help in increasing their Seebeck coefficients and thereby have applications in thermoelectric devices. For example, FeSb$_2$ shows the highest thermoelectric power of -45000 $\mu$V/K. \cite{Bentien}. RuGa$_3$, OsGa$_3$, FeGa$_3$ also show large Seebeck coefficient of -274, -577 and -563 $\mu V$/K respectively. \cite{Amagai}

Among these transition metal based intermetallic semiconductors, FeGa$_3$ is one of the most well studied materials. It has a tetragonal crystal structure and it belongs to $P4_2/mnm$ space group symmetry. \cite{Amagai, Umeo, Arita1, Debashis1} The unit cell has four Fe atoms, arranged in two pairs of dimers. The Fe-Fe distance in the dimer is 2.77 \AA{}, and these dimers are separated from each other by 6.55 \AA{} along $c$ direction. and are surrounded by Ga atoms. Different indirect measurements, such as temperature dependent resistivity, temperature dependent magnetic susceptibility, a combination of photoelectron spectroscopy (PES) and inverse photoelectron spectroscopy (IPES) measurements, have been carried out to estimate the bandgap of this material, \cite{Amagai, Umeo, Tsujii, Arita1, Hadano, Debashis1} and is found to be between 0.3 - 0.5 eV. Local moment on the Fe atom in FeGa$_3$ has been probed by Mossbauer spectroscopy. It shows no detectable local moment on the Fe atoms in this system. \cite{Tsujii, Dezsi,Debashis1}. However, in a recent neutron diffraction measurement on FeGa$_3$, the possibility of the existence of a magnetic moment on the Fe atoms with an incommensurate canted antiferromagnetic arrangement of the magnetic moments has been indicated.\cite{Gamza} Muon spin resonance measurements have also indicated the presence of a magnetic moment on the Fe atoms in FeGa$_3$. \cite{Storchak} %The presence of magnetic moment on the Fe atoms, which are in the form of nearly isolated dimers, is an interesting point as it opens up the possibility for the observation of quantum entanglement of this system and its possibility to be used for quantum computing.

It has been observed that first-principles based electronic structure calculations with local density approximation (LDA) and generalized gradient approximation (GGA) exchange-correlation functionals, reproduce the bandgap similar to experimental values with a non-magnetic state of FeGa$_3$ where the Fe atoms carry no magnetic moment. \cite{Ulrich2, Imai, Yin, Osorio} A close agreement between the experimental and calculated (using LDA and GGA approximation) values of band gap and reported nonmagnetic state of Fe atom in FeGa$_3$ suggests that the LDA and GGA approximations are likely to be sufficient and an on-site Coulomb repulsion term (U), which is important in several 3d transition metal based compounds and intermetallics, may not be significant in this system under ambient conditions.

Attempts to explain the recent reports regarding the observation of magnetic moment \cite{Gamza, Storchak} on Fe atoms have been made using the density functional theory (DFT)+U approach, where the presence of on-site Coulomb repulsion is incorporated in the first principles calculation as an additional parameter. In the local spin density approximation (LSDA) + U calculations reported by Yin \textit{et al.}, both ferromagnetic and antiferromagnetic arrangements of the moments on the Fe atoms in the dimers were considered. It is reported that by starting the calculations with a ferromagnetic arrangement of the moments on the Fe atoms, irrespective of the value of U, the minimum energy state of the system is ferromagnetic with no bandgap. In the case of an initial antiferromagnetic arrangement of the moments on the Fe atoms in the dimer, for U $\geq$ 3.5 eV, the minimum energy state of the system is antiferromagnetic without a bandgap. For smaller values of U ( 2 eV to 3  eV), the minimum energy state has an antiferromagnetic configuration with a bandgap (0.52 eV - 0.13 eV), and for U $\leq$1.5 eV, the system is non-magnetic with a bandgap close to the experimental value. Arita et al. report the band dispersion for calculations with finite U ( U = 3 eV) without the mention of the nature of the alignment of the magnetic moments on the Fe atoms in the dimer. They have compared the calculations with the measured ARPES data for FeGa$_3$ and a relatively better matching is reported for U $\sim$ 3 eV as compared to the calculations with U = 0 eV. 

Based on the above mentioned results reported in the literature, some important questions regarding the nature of FeGa$_3$ arise, namely the nature of the magnetic state of FeGa$_3$, and whether the Fe atoms carry a magnetic moment. Further, if FeGa$_3$ is indeed diamagnetic, then how do we reconcile the fact that a magnetic moment is observed in the neutron diffraction and muon spin resonance experiments. In case the Fe atoms in FeGa$_3$ carry a magnetic moment, the nature of the alignment of the magnetic moments on the Fe atoms is to be ascertained.

At this point, it is also important to draw our attention to the point that in several reports, where Fe or the Ga atoms in FeGa$_3$ have been substituted with an electron or a hole donor, a magnetic moment is observed and the system shows a ferromagnetic alignment of magnetic moments \cite{Umeo, Gamza} Both electron (by Ge), and hole dopings (by Zn) at Ga site render the system magnetic.\cite{Umeo,Gamza,Bittar,Gippius} Doping at Fe site by Mn also makes the system magnetic.\cite{Gamza} A magnetic moment of $0.155$ $\mu_B$ has been found as a result of $5\%$ Co doping by Bittar \textit{et. al.} and Gippius \textit{et. al.}\cite{Bittar,Gippius}. These observations indicate the importance of disturbing the Fe dimers to create a magnetic moment. In all the references mentioned above, the Fe dimers have been disturbed by a non-isoelectronic dopant in the Ga and Fe site. The other way to disturb the Fe dimer is to have an FeGa$_3$ sample that has a large number of defects, dislocations, grain boundaries etc, all of which are also expected to disturb the Fe dimer. It is to be noted that these factors are expected to have a considerably less influence as compared to the effect of  doping on the Fe and/or Ga sites.

In this work, we carry out photoelectron spectroscopy (PES), magnetic measurements (M-H and temperature dependent magnetization measurement) and X-ray absorption near edge spectroscopy (XANES) experiment on single crystal FeGa$_3$ samples to find the answer of the questions mentioned above regarding the magnetic properties of FeGa$_3$. Further, to understand the effect of extrinsic factors like grain boundaries, disorder, defects etc, the same measurements have also been carried out on polycrystalline FeGa$_3$ sample. The polycrystalline sample is expected to have crystal defects in a significantly larger amount as compared to the single crystal. We find that the experimentally observed valence band spectrum matches with the results of ab-initio calculation with U = 0 to 1 eV very well. We also find that FeGa$_3$ is inherently diamagnetic in nature and our studies indicate that the magnetic moment reported in FeGa$_3$ is basically due to non idealities i.e. the presence of disorder, defects, grain boundaries etc.

\section{Experimental and Computational details}

FeGa$_3$ single crystals have been grown using Ga flux method. The details of sample preparation technique are described in our earlier communication. \cite{Debashis2} The chemical composition of the single crystal was checked using EDAX, which shows a Fe and Ga ratio of 26:74. For structural information, X-ray diffraction (XRD) measurement has been carried out using a Bruker D8 discover system with Cu $k_\alpha$ source. A small portion of the single crystal was powdered for this purpose. We have carried out Rietveld refinement of the XRD data to find out its crystal structure. X-ray near edge spectroscopy (XANES) measurements have been performed at the EXAFS beamline (BL-09), Indus-2 synchrotron radiation source, RRCAT, Indore. The Fe K-edge absorption data have been measured in fluorescence mode using a Vortex detector where the energy has been calibrated using a Fe foil.

\begin{figure}[h]
\centering
\includegraphics[width=0.45 \textwidth]{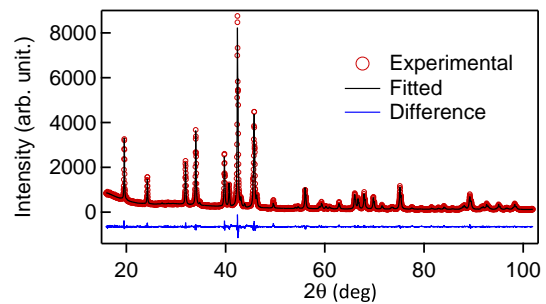}
\caption{ X-ray diffraction pattern of FeGa$_3$ powder at ambient pressure using Cu $K_{\alpha}$ source.}
\end{figure}

PES experiments were carried out using SPECS Phiobos 150 analyser. The He-I line (21.2 eV photon) was used to probe the valence band in our experiment with an energy resolution of $\sim$ 15 meV. The base pressure during the measurement was $\sim7~ \times 10^{-11}$ mbar. X-ray photoelectron spectroscopy (XPS) measurements were performed in the same system using a Mg $K_\alpha$ source. The sample was fractured and absence of surface contaminants was confirmed by the absence of Carbon and Oxygen $1s$ peak at ~ 287 eV and ~ 531 eV binding energy, respectively, before the measurements. The Fermi edge of the sample was aligned to Fermi edge of Au sample for the same analyzer settings.

M-H measurements have been carried out on FeGa$_3$ single crystal and polycrystalline samples using a Quantum design SQUID magnetometer at 2 K and 300 K upto a magnetic field of 7 Tesla. Temperature dependent magnetization measurements at different applied magnetic fields for these two samples were also carried out.

\begin{figure}[h]
\centering
\includegraphics[width=0.4 \textwidth]{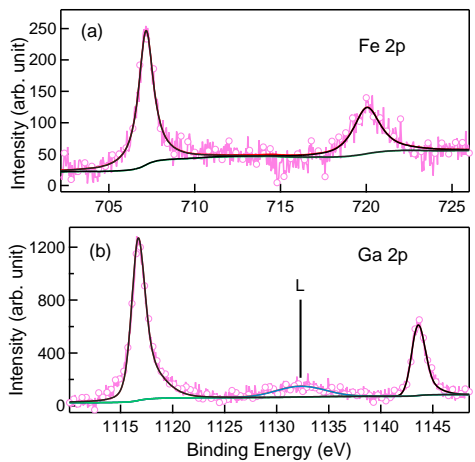}
\caption{X-ray Photo-emission spectra of FeGa$_3$ recorded at 15K using Mg K$\alpha$ photon source. In fig. 2 (b), 'L' denotes the loss feature in the photo-emission spectrum.}
\end{figure}

To understand the experimental results, we have calculated the electronic structure of FeGa$_3$ by DFT method. Vienna Ab-initio Simulation Package (VASP), based on projector augmented wave (PAW) has been used for this purpose. \cite{Kresse1, Kresse2}  We have considered generalized gradient approximation (GGA)  for exchange-correlation functional, given by Perdew, Burke and Ernzerhof (PBE) \cite{Perdew} in our calculations. The plane wave cut-off energy for the basis set 400 eV. We have used Monkhorst-Pack scheme with $k$-meshes of $11\times 11\times 10$ for Brillouin zone integration and the energy convergence criterion in SCF cycles was chosen to be $10^{-6}$ eV. The geometric structure of FeGa$_3$ has been optimized by minimizing the forces on individual atoms by applying the condition that the total force on each atom should be below 10$^{-2}$ eV/\AA{}. We have also carried out DFT +U calculation to find out the effects of on-site Coulomb correlation in the Fe 3d electrons by employing GGA + U approach with U  varying from 1 to 5 eV.

\section{Results and discussion}

\subsection{Structural analysis}
Room temperature XRD pattern of FeGa$_3$ is shown in Fig 1. The results obtained from Rietveld refinement of the XRD pattern confirms the tetragonal structure of the material with $ P4_2/mnm$ space group symmetry. The lattice parameter values obtained are {\it a} = 6.267 ($\pm$0.00016) \AA{} and {\it c} = 6.560 ($\pm$0.00019) \AA{}. These values are in close agreement with the data reported in literature. \cite{Amagai, Umeo, Tsujii, Arita1, Hadano,Ulrich2, Debashis1} The fractional coordinates of the atoms within the unit cell are shown in Table - I. A comparison of the lattice parameters and fractional coordinates of this single crystal FeGa$_3$ with polycrystalline FeGa$_3$ reported earlier by us and by others \cite{Ulrich2,Debashis1} clearly shows that these samples are crystallographically identical.

\begin{center}
\begin{table*}
\caption{Rielveled refinement result for FeGa$_3$.}
%\begin{tabular}{m{1cm} m{1.5cm} m{1cm} m{1cm} m{1cm} m{1cm}}
 \begin{tabular}{P{1cm}P{1.8cm}P{1.8cm}P{1.8cm}P{1.8cm}P{1cm}}
\hline
$Atom$ &	$Position$ &		$x$ &	$y$ &	$z$ &	$occ$\\
\hline

Fe&			4f&		$0.34412(29)$&	$0.34412(29)$&	$0.00000$&	$1.00$\\
Ga1&		4c&		$0.00000$&	$0.50000$&	$0.00000$&	$1.00$\\
Ga2&		8j&		$0.15593(15)$&	$0.15593(15)$&	$0.26037(22)$&	$2.00$\\
\hline
\end{tabular}
\end{table*}
\end{center}

 \begin{figure}[h]
 \centering
 \includegraphics[width=0.4 \textwidth]{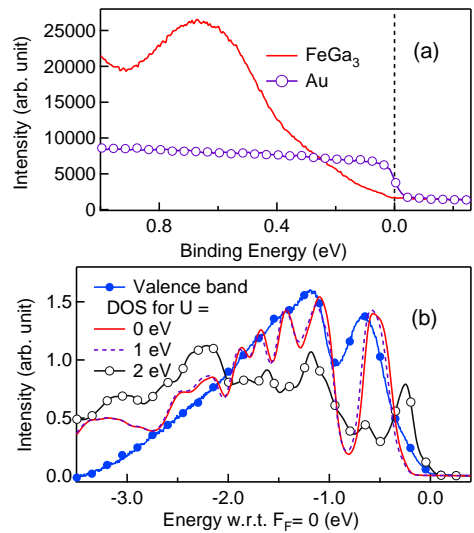}
 \caption{(a) Experimental valence band of FeGa$_3$ and Au have been plotted in the same analyser setting. (b) Experimental valence band spectrum has been compared with the DOS obtained from ab-initio calculation with GGA+U approximation, for U = 0, 1 and 2 eV.}
 \end{figure}

\subsection{Photo-electron spectroscopy}

The XPS core level spectra of Fe $2p$ and Ga $2p$ are shown in fig. 2 (a) and fig. 2 (b) respectively. In case of Ga $2p$, a loss feature has been found at $\sim$ 1132 eV, which is in good agreement with the loss feature observed in case of elemental Ga at $\sim$ 1130 eV. \cite{Pes} We have calculated the surface composition of FeGa$_3$ from the intensity ratio of photo emission spectra of Fe $2p$ and Ga $2p$, which is found to be 24$\%$ and 76$\%$ for Fe and Ga respectively. This is nearly same as the bulk composition calculated from EDAX measurement as mentioned before. The binding energy of Fe $2p_{3/2}$ and Ga $2p_{3/2}$ peak in FeGa$_3$ has been found to be 707.15 eV and 1116.5 eV, respectively, whereas in case of elemental Fe and Ga, the binding energy values are 706.8 eV and 1116.4 eV respectively. These small differences in the corresponding binding energies are possibly due to the hybridization between the valence bands of the parent elements. The valence band of FeGa$_3$ obtained from UPS measurement is shown in fig 3. In fig. 3 (a), we have also plotted the valence band of gold recorded at 10 K in the same analyzer settings for comparison. It is observed that the density of states near the Fermi level in FeGa$_3$ is negligible. This confirms that the FeGa$_3$ single crystal is a semiconductor. Our temperature dependent resistivity measurements on polycrystalline FeGa$_3$ samples also confirm the semiconducting nature of the compound. \cite{Debashis1}

\begin{figure}[h]
\centering
\includegraphics[width=0.48 \textwidth]{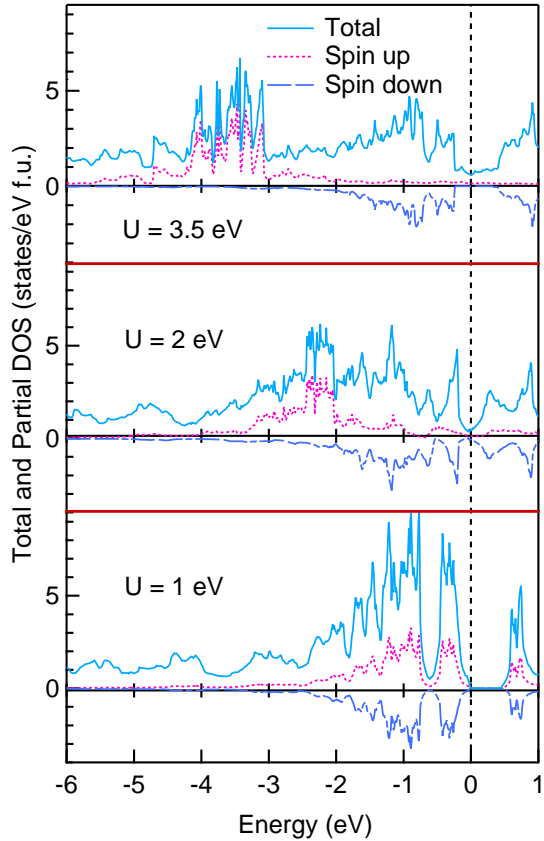}
\caption {Total and Spin projected DOS of Fe 3d states obtained from ab-initio calculation with GGA+U approximation, for U = 1, 2 and 3.5 eV are shown. }
\end{figure}

 \subsection{Ab-initio calculations}
 
To understand the experimental valence band spectrum observed by PES, we have carried out DFT based electronic structure calculation with GGA approximation. We find a bandgap of $\sim$ 0.43 eV from our calculations. This value is similar to those reported earlier. \cite{Arita1, Ulrich2, Imai, Yin, Osorio, Anita} Figure 3(b) shows the experimental  valence band spectrum and density of states obtained from ab-initio calculation with U = 0, 1 and 2 eV for 3d orbital of Fe atom, it is observed that starting with either the ferromagnetic or the antiferromagnetic alignment of the Fe moments, the ground state after energy minimization is observed to be same for U = 0 and U = 1 eV i.e. a semiconductor. For the sake of comparison of calculated DOS with experimentally observed valence band, we have broadened and shifted the calculated DOS. Similar method has been followed earlier by Arita \textit{et. al.} \cite{Arita1} A good match with all the  valence band features including the dip at $\sim$ -0.88 eV is observed for the U=0 and U=1 eV calculations. The slight mismatch between the calculated and experimental DOS curve close to E = 0, is attributed the presence of band tailing (Urbach tail), which is expected due to the presence of small quantity of crystalline defects.  However, the density of states calculated with U = 2 eV, is significantly different from the experimentally observed valence band and we find a very small density of states at the Fermi level. Further, this feature is prominent in fig. 4, where the total as well as 3d spin up and spin down partial DOS of Fe atom are shown. We find from fig. 4 that, as U increases beyond 1 eV, the DOS of 3d spin up and spin down changes and consequently the system becomes magnetic. In the d-p hybridization induced semiconductor FeGa$_3$, \cite{Ulrich2, Amagai, Yin} the bandgap is formed due to hybridization between the 3d band of Fe atom and 4p band of Ga atom. In case of FeGa$_3$, the applied on-site Coulomb repulsion U, pulls the Fe 3d band away from the Fermi level, which results in a weakening of the Fe3d-Ga4p hybridization. Consequently, there is an increase in the DOS corresponding to Ga 4p states at the Fermi level moving the system towards metallic. \cite{Debashis2} As the value of U increases to 3.5 eV, (shown in fig. 4) the system definitely becomes metallic. From PES measurements we find that, FeGa$_3$ is a semiconductor at ambient condition. So, the results of our ab-initio calculations suggest the value of U to be below 2 eV, and in this range, FeGa$_3$ has been found to be non magnetic. Our earlier work on FeGa$_3$ at high pressure also supports this conclusion on the value of U. \cite{Debashis2}

 \begin{figure}[h]
 \centering
 \includegraphics[width=0.45 \textwidth]{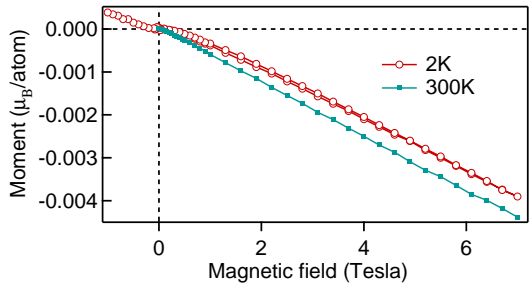}
 \caption{M-H curve for FeGa$_3$ single crystal at 2 K and 300 K.}
 \end{figure}

\subsection{Magnetic measurements}

To ascertain and analyze the magnetic properties of FeGa$_3$, we have carried out detailed magnetic measurements. The results of the magnetization measurements on FeGa$_3$ single crystal are shown in fig. 5. M-H measurement on the single crystal of FeGa$_3$ at 300 K clearly shows a diamagnetic behaviour, although at 2 K, a nonlinearity is observed in the magnetization curve around an applied magnetic field of 0.5 Tesla. To confirm the diamagnetic property of FeGa$_3$ single crystal, temperature dependent magnetization measurements have been carried out at various applied magnetic field (0.2 T, 0.5 T and 2 T) in the temperature range of 2 K to 300 K. It may be noted that, in case of FeGa$_3$ single crystal, we could not collect the magnetization data below the applied field of 0.2 T, due to the very low value ($\sim$ 10$^{-5}$ emu) of magnetic moment. Figure 6 shows the variation of magnetic moment with respect to temperature. We find that the value of magnetic moment is negative in all the cases and as the applied field increases, the value of the magnetic moment becomes more negative. For temperatures beyond 100 K, the magnetization is almost constant with temperature for all the applied magnetic fields. We also find that at low temperature (below 50 K), there is a distinct decrease in the value of magnetization with temperature, indicating a small paramagnetic background along with the dominant diamagnetic nature.

\begin{figure}[h]
\centering
\includegraphics[width=0.45 \textwidth]{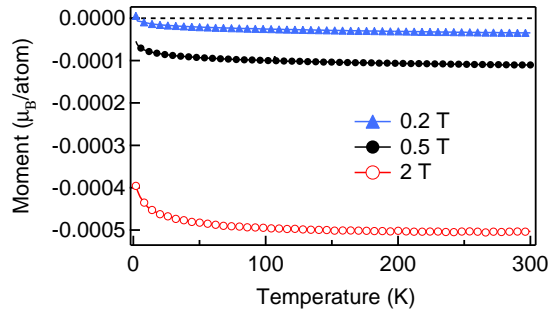}
\caption{Variation of magnetic moment with temperature in the range of 300 K to 2 K at different applied magnetic field for FeGa$_3$  single crystal}
\end{figure}

The M-H curve for polycrystalline FeGa$_3$ at 2 K and 300 K are shown in fig. 7. At 300 K, for very low applied fields, there is an increase in the magnetic moment upto $\sim$ 0.5 T, which is followed by a continuous decrease in the moment. At $\sim$ 1.6 T, the magnetic moment is very close to zero and remains negative thereafter, thereby showing a dominant diamagnetic behavior.

From the M-H data obtained at 2 K, we find that at very low applied fields upto $\sim$ 0.04 T, the magnetization is negative and decreases with increase in field. With increase in the applied field upto $\sim$ 1.5 T, the magnetization increases and becomes positive. With a further increase in the applied field upto 7 T (which is the highest field studied in this work), there is a continuous decrease in the magnetization, although, the magnetization remains positive throughout this range. This non-monotonic change in the magnetization with applied field is a clear indication of the presence of simultaneous competing processes in the system. To have a better understanding of these competing processes, the magnetization vs temperature measurements at different applied fields (from 0.05 T to 0.5 T) were carried out and are shown in fig. 8.

\begin{figure}
\centering
\includegraphics[width=0.45 \textwidth]{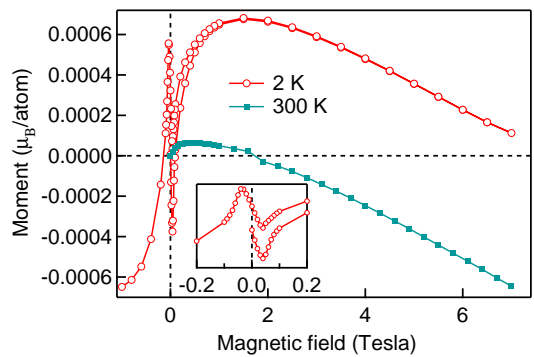}
\caption{M-H curve for FeGa$_3$ polycrystalline sample at 2 K and 300 K. Inset shows the zoomed portion of the 2 K data.}
\end{figure}

The M-T data for a low applied field of 0.05 T shows a non-monotonic behaviour. At the lowest temperature of 2 K, the magnetization is negative and it increases with increase in temperature. The magnetic moment crosses the temperature axis at $\sim$ 4.7 K and has a maximum value at $\sim$ 13.2 K. Beyond this temperature, there is a monotonic fall in the value of the magnetic moment with increase in temperature, but it remains positive throughout the temperature range studied in this work. In case of M-T curve for higher applied fields, the high temperature data is qualitatively similar to the data observed for low applied fields in the same temperature range. However at  very low temperatures, the non-monotonicity observed at low field, is not seen in the data measured at high field. The increase in magnetization with increasing temperature in the low temperature region at low field (0.05 T and 0.1 T) is attributed to an antiferromagnetic ordering in the material. Gamza \textit{et. al.} have also observed an antiferromagnetic ordering in the neutron diffraction data at 2 K, which was absent at 300 K. From the M-T data at high temperatures, we find that the magnetisation is positive at all the applied fields, indicating that the paramagnetic component dominates. Combining this observation with the shape of the M-H curve, where a decrease in M with applied field H is observed  (fig 7), we can unambiguously say that, both the diamagnetic and the paramagnetic components coexist in this powdered FeGa$_3$ sample.

\begin{figure}
\centering
\includegraphics[width=0.45 \textwidth]{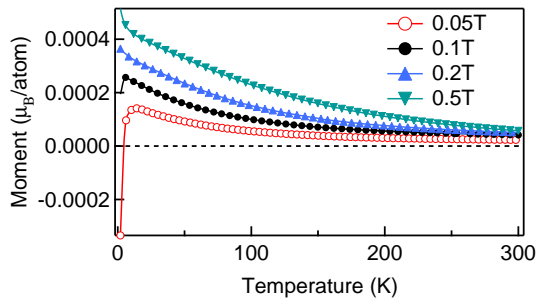}
\caption{Variation of magnetic moment with temperature in the range of 300 K to 2 K at different applied magnetic field for polycrystalline FeGa$_3$. }
\end{figure}

In summary, we find significant differences in the magnetic properties between single crystal and polycrystalline FeGa$_3$. First, M-T data of FeGa$_3$ single crystal show almost a negative constant value of the magnetization with temperature, which clearly indicates a diamagnetic behavior of the sample. However, in case of polycrystalline FeGa$_3$, a paramagnetic type variation of the magnetization has been observed in the M-T curve with a positive value of magnetization at the highest magnetic field. Second, the value of the magnitude of the magnetic moment of polycrystalline FeGa$_3$ at 7 T in 300 K (-6.5$\times 10^{-4} \mu_B / atom$) is one order of magnitude smaller, as compared to the magnetic moment of FeGa$_3$ single crystal (-4.4$\times 10^{-3} \mu_B / atom$). With lowering of temperature, the ordering of these paramagnetically ordered moments increase, which results in a further increase in the moment (see M-H data at 2 K of polycrystalline FeGa$_3$).

The results of the magnetic measurements on single crystal and polycrystalline FeGa$_3$ coupled with the PES measurements and the first principles calculations, provide a complete and consistent information about the electronic structure of FeGa$_3$. We find that, FeGa$_3$ single crystal is primarily diamagnetic in nature. As defects, disorder, grain boundaries are produced, as in the polycrystalline powdered material, the Fe dimers are broken into individual Fe atoms, which contribute in the paramagnetically ordered moment of the polycrystalline sample. Thus in the polycrystalline sample, we get the resultant of the magnetic moments from the negative diamagnetic component and positive paramagnetic component. Hence the moment of polycrystalline FeGa$_3$ sample is lower than the moment of single crystal FeGa$_3$. So, the magnetic moment on the Fe atoms in FeGa$_3$ is not an inherent property of FeGa$_3$, rather, an effect of disturbance of the Fe pairs in FeGa$_3$. It may be noted that the presence of grain boundaries and defects in general goes undetected in a powder XRD measurement. However, extremely small magnetic moments that may be generated on a few disturbed Fe dimers in bulk and the surface FeGa$_3$, which will result in an observed magnetic moment over the weak diamagnetic background. Thus introduction of on-site Coulomb repulsion (either through DFT + U or DMFT) on the Fe 3d electrons to explain the magnetic moment on Fe 3d electrons, electronic structure calculations at ambient pressure, is not necessary in FeGa$_3$, as is clear from our detailed PES and magnetic measurements.

\begin{figure}
\centering
\includegraphics[width=0.45 \textwidth]{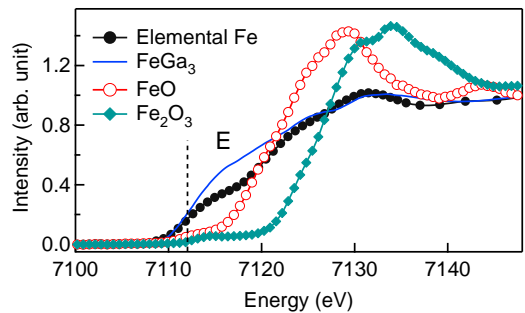}
\caption{XANES measurement at Fe edge of FeGa$_3$, Fe metal foil, Fe$^{2+}$ and Fe$^{3+}$ compound. The energy range were calibrated with Fe metal foil.}
\end{figure}

It is to be noted here that, it has been shown in the literature to exist a complex mixed charge state in Fe atom in FeGa$_3$ \cite{Gamza, Anita} Therefore, we have determined the charge state of Fe atoms in FeGa$_3$ using XANES measurements. Figure 9 shows the XANES data recorded at the Fe edge in FeGa$_3$, along with Fe absorption edge in Fe metal foil, FeO and Fe$_2$O$_3$. The XANES spectra of FeGa$_3$ looks similar to the Fe metal foil and it differs significantly from Fe$^{2+}$ and  Fe$^{3+}$ oxidization state. Thus, result of our XANES measurement, rules out the possibility of the presence of a charge state on Fe atom in FeGa$_3$. This observation further supports our conclusion that invoking an on-site Coulomb repulsion for Fe atom through LDA+U or DMFT calculation is not required to determine various properties of FeGa$_3$ at ambient pressure.

\section{Conclusion}

In this work we revisit the electronic and magnetic structure of FeGa$_3$ using XRD, PES, XANES, magnetic measurements and ab-initio calculations to understand and explain the results of our experiments vis-a-vis those reported in the literature. We find that single crystal FeGa$_3$ sample is inherently diamagnetic in nature. Magnetic moment and its manifestation in the magnetic measurements is observed in the polycrystalline samples only.  Thus, the on-site Coulomb repulsion on the Fe 3d electrons, which was introduced in the electronic structure calculation of FeGa$_3$ to explain the presence of the magnetic moment in this system at ambient condition, is not warranted. The small magnetic moment found in different measurements on polycrystalline FeGa$_3$ samples is primarily attributed to the breakdown of symmetry around the Fe dimers and/or breaking of the Fe dimers at the grain boundaries, defects etc. This results in a contribution in the paramagnetically ordered magnetic moments in the polycrystalline sample along with the intrinsic diamagnetic contribution of perfectly ordered FeGa$_3$. This aspect is clearly brought out in our magnetisation data, where we have compared the magnetic measurements in single crystal and polycrystalline samples. Our combined results of experiments and ab-initio calculations  consistently help in resolving the magnetic nature of FeGa$_3$.

\section{Acknowledgement}

The authors thank Shri Ajay Kak for providing experimental support. Mr. Kranti Kumar Sharma is thanked for supporting in magnetic measurements. Dr. P. A. Naik is cordially thanked for encouragement. CK and AC thank Dr. A. Banerjee and Scientific Computing Group, Computer Division, RRCAT for their support. DM also would like to thank RRCAT for financial support.

\end{document}